\documentstyle[prd,aps,floats]{revtex}

\begin{document}
\preprint{SUSSEX-AST 97/8-2, astro-ph/9708247}
\draft

%
%
\input epsf
\renewcommand{\topfraction}{1.0}

\title{Perturbation evolution in cosmologies with a decaying cosmological 
constant} 
\author{Pedro T.~P.~Viana and Andrew R.~Liddle}
\address{Astronomy Centre, University of Sussex, Falmer, Brighton BN1
9QH,~~~U.~K.}  
\date{\today} 
\maketitle
\begin{abstract}
Structure formation models with a cosmological constant are successful
in explaining large-scale structure data, but are threatened by the
magnitude--redshift relation for Type Ia supernovae. This has led to
discussion of models where the cosmological `constant' decays with
time, which might anyway be better motivated in a particle physics
context. The simplest such models are based on scalar fields, and general 
covariance demands that a time-evolving scalar field also supports spatial 
perturbations. We consider the effect of such perturbations on the growth of
adiabatic energy density perturbations in a cold dark matter component. We 
study two types of model, one based on an exponential potential for the 
scalar field and the other on a pseudo-Nambu Goldstone boson. For each 
potential, we study two different scenarios, one where the scalar field 
presently behaves as a decaying cosmological constant and one where it 
behaves as dust. The initial scalar field perturbations are fixed by the 
adiabatic condition, as expected from the inflationary cosmology, though in 
fact we show that the choice of initial condition is of little importance. 
Calculations are carried out in both the zero-shear (conformal newtonian) 
and uniform-curvature gauges. We find that both potentials allow models 
which can provide a successful alternative to cosmological constant models. 
\end{abstract}

\pacs{PACS numbers: 98.65.Dx, 04.40.-b \hspace*{0.8cm} Sussex preprint 
SUSSEX-AST 97/8-2, astro-ph/9708247}


\section{Introduction}

Ever since the demise of the Standard Cold Dark Matter model, which
proved unable to simultaneously match the COBE observations of cosmic
microwave background anisotropies, the abundance of rich galaxy
clusters and the shape of the galaxy correlation function, it has been
fashionable to study variations on this basic theme. A popular
alternative is to lower the matter density, thus shifting the epoch of
matter--radiation equality in a desirable direction. To keep the
possibility that standard models of inflation, generating a
spatially-flat Universe, might be responsible for the seed
perturbations, the most popular version of this model is to introduce
a cosmological constant $\Lambda$ to maintain the spatial flatness.
This has proven a compelling framework within which it is possible to
understand the formation and evolution of large-scale structure in the
Universe \cite{LamCos,LLVW}.

The favoured models, until recently, had the energy density
$\Omega_{\Lambda}$ in the cosmological constant lying in the range 0.5
to 0.7. As well as giving a good fit to large-scale structure
observations, these models have support from two further sources. If
one assumes the baryon density in the Universe is that predicted by
standard big bang nucleosynthesis \cite{SBBN}, and that large galaxy
clusters provide a representative sample of the universal ratio
between baryons and the total amount of matter in the
Universe~\cite{ClusBar}, then the matter density must be significantly
below one. Further, the need to have the age of the Universe, $t_{0}$,
exceeding the age of the globular clusters in our galaxy suggests that
$\Omega_{\Lambda}$ should be as large as possible. Until recently the
required value for $t_{0}$ was usually around 14 Gyrs
\cite{AgeofUni}, but a revised Cepheid distance scale due to new distance 
estimates by the satellite Hipparcos has brought this down to
something more like 12 Gyrs \cite{HippRes,HippAge}. Given that most
measurements (even taking into account Hipparcos's revision of the
Cepheid distance scale) seem to suggest that the present value of the
Hubble parameter, $h$, is at least 0.6 (in the usual units of $100
\;{\rm km}\,{\rm s}^{-1}\,{\rm Mpc}^{-1}$)
\cite{Hubble06} and perhaps even larger \cite{Hubblelarger}, in a flat 
Universe we would then need $\Omega_{\Lambda}>0.3$ (0.55 if
$t_{0}>14\,{\rm Gyrs}$).

Unfortunately, these models have been dealt a serious blow by the
preliminary results from the Supernova Cosmology Project \cite{Perl},
which attempts to determine the magnitude--redshift diagram of Type Ia
supernovae.  They place a 95 per cent confidence upper limit of
$\Omega_{\Lambda} < 0.5$ for flat Universes. This is significantly
stronger than earlier limits from the galaxy velocity distribution
\cite{GalVel}, galaxy outflows from voids
\cite{DekelR} and the statistical analysis of the frequency of gravitational 
lensing of high-redshift quasars \cite{K96}, all coming in around
$\Omega_{\Lambda} < 0.7$. While the $\Lambda$CDM model remains viable
with these smaller $\Omega_{\Lambda}$ values, this is seen as much
less attractive because, as in the critical-density case, the required
matter density is well above that given by direct observation.

A way out of this dilemma is to move to cosmological models where the
cosmological constant is substituted by a dynamical quantity which
decays with time 
\cite{Arblam,FAFM,Wett88,Wett95,FCAI,RatPee,Weiss,CDFri,FerrJoy,CDS}. 
Within these models it is possible to relax the
constraints resulting both from the frequency of gravitational lensing
of high-redshift quasars and from Type
Ia supernovae \cite{RatQui,FCAI,SteinNat,CDFri}. While in some ways this is
clearly a regressive step, introducing more freedom into the model, it
can also be argued that such a situation may be more natural on
particle physics grounds. For example, there is the well-known
difficulty within quantum field theory to understand the very small
vacuum energy density, $\mu_{\rm vac}=(0.003\,{\rm
eV})^{4}\Omega_{\Lambda}$, required by a cosmological constant. If not
strictly zero, due to some yet unknown cancellation mechanism, one
would expect $\mu_{\rm vac}$ to be between 50 and 120 orders of
magnitude larger \cite{CCcal}! A decaying cosmological constant term
would be a simple way of reconciling a very large vacuum energy
density early on in the Universe with an extremely small one at
present.

Some authors have simply assumed more or less {\em ad hoc} decay laws
for the cosmological constant term \cite{Arblam}. 
By comparing predictions of the models with observations it was then hoped 
that the correct decay law could be recovered, which would then shed some 
light on the possible physical process behind the decaying cosmological 
constant term. However, a credible mechanism for obtaining such a term 
already exists, which is to assume the existence of a scalar field 
presently relaxing towards the minimum of its potential 
(see e.g.~\cite{FAFM,Wett88,Wett95,FCAI,RatPee,Weiss,CDFri,FerrJoy,CDS}). 
Scalar fields are not only predicted to exist by 
some particle physics theories that go beyond the Standard Model, but are
also the most plausible engine behind a possible inflationary period
in the very early Universe \cite{Inf}. The overall dynamics of the
Universe in the presence of a relaxing scalar field, and its
consequences for several classical cosmological tests, has been
studied in detail by various authors \cite{RatQui,ObsTests}.

Our aim in this paper is to study the effect of spatial perturbations in the
cosmological `constant' term given by such a field, in particular on the 
growth of the
matter perturbations. When one has the standard constant $\Lambda$
term, there is no possibility of any perturbations in it. However, as
soon as one permits any form of time variation, general covariance
immediately implies that it must be able to support spatial
perturbations. Despite this, presumably because of the development of
this line of research from the original constant case, with few
exceptions \cite{RatPee,Weiss,CDFri,FerrJoy,CDS} most authors have not
looked at the possible effect of spatial perturbations in the
background value of the scalar field on the growth of perturbations in
the matter distribution.

\section{Equations and initial conditions}

\label{incon}

We will assume that the background space-time contains an ideal fluid
and a scalar field. The equation of state of the ideal fluid, relating
its background pressure, $p_{\gamma}$, to its background energy
density, $\mu_{\gamma}$, is $p_{\gamma}=(\gamma-1)\mu_{\gamma}$, where
$\gamma$ is a constant. The background energy density and pressure
associated with a minimally coupled real scalar field with potential
$V(\phi)$ are given by
\begin{equation}
\label{sfpe}
\mu_{\phi}=\frac{1}{2}\,\dot{\phi}^{2}+V(\phi)\quad ; \quad
p_{\phi}=\frac{1}{2}\,\dot{\phi}^{2}-V(\phi)\,.
\end{equation}
The overdots represent derivatives with respect to coordinate time $t$.

The evolution of these background quantities is described by the
Friedmann equation,
\begin{equation}
\label{Friedmann}
H^{2}=\frac{8\pi G}{3}\left[\mu_{\gamma}+\frac{1}{2}\,\dot{\phi}^{2}
	+V(\phi)\right]\,,
\end{equation}
together with the two energy conservation equations 
\begin{eqnarray}
\label{enconsfl}
\dot{\mu}_{\gamma}=-3H\gamma\mu_{\gamma}\,,\\
\label{enconssf}
\ddot{\phi}+3H\dot{\phi}+V,_{\phi}=0\,,
\end{eqnarray}
for the ideal fluid and the scalar field respectively. Here
$H\equiv\dot{a}/a$ is the Hubble parameter, $a$ the cosmic scale
factor, and $,_{\phi}$ represents a derivative with respect to $\phi$.

We shall carry out a fully relativistic treatment of the
perturbations, using a formalism based on a series of papers by Hwang
\cite{Hw1,Hw2,Hw3,Hw4,Hwetal}. As always, the equations describing 
the evolution of the perturbations are very complicated and we have 
relegated their discussion to our two Appendices. A crucial question when 
studying perturbations is the choice of gauge. We shall consider two 
choices. Mainly we shall use the zero-shear gauge (ZSG) [also known as 
conformal newtonian or longitudinal], but we shall also solve the equations 
in the uniform-curvature gauge (UCG) as a check on the accuracy of the 
numerical calculations.

The perturbation equations, described in the Appendices, relate the
perturbed part of the metric variables, $\alpha$ (perturbed part of
the lapse function), $\varphi$ (perturbed part of the spatial
curvature), $\chi$ (perturbed part of the shear) and $\kappa$
(perturbed part of the expansion scalar), to the perturbed part of the
matter variables, $\epsilon=\epsilon_{\gamma}+\epsilon_{\phi}$
(perturbed part of the total energy density),
$\varpi=\varpi_{\gamma}+\varpi_{\phi}$ (perturbed part of the total
pressure) and $\Psi=\Psi_{\gamma}+\Psi_{\phi}$ (perturbed part of the
total energy density flux, or total fluid four-velocity, depending on
the frame chosen). In the ZSG $\chi$ is chosen to vanish, while in the UCG
$\varphi$ vanishes. In all, we will need to numerically
integrate a system of seven simultaneous first-order ordinary differential
equations, formed by the background equations, Eqs.~(\ref{back1}) to
(\ref{back3}), where Eq.~(\ref{energyevol}) gives $\mu_{\gamma}$, and
the four perturbation equations resulting from either
Eqs.~(\ref{pertZSG7}) and (\ref{pertZSG8}) (in the ZSG), or
Eqs.~(\ref{pertUCG7}) and (\ref{pertUCG8}) (in the UCG).

We must now specify the initial conditions required in order to solve
this system. We shall do that by providing the initial values for the
quantities $\epsilon_{\gamma}$, $\Psi_{\gamma}$, $\delta\phi$ and
$\delta{\phi}'$. The first two give the initial values for $\varphi$
and $\varphi'$ (ZSG) ($\chi$ and $\chi'$ in the UCG) by means of
Eqs.~(\ref{pertZSG3}) and (\ref{pertZSG4}) [Eqs.~(\ref{pertUCG3}) and
(\ref{pertUCG4}) in the UCG].

We make two requirements on the initial conditions. The first is to
choose the initial value of $\Psi_{\gamma}$ so that only the growing
mode of the solution for the evolution of an ideal fluid in an
Einstein-de Sitter Universe is present (see the exact analytical
solutions in Table 1 of Ref.~\cite{Hw2}), as this is the situation we
would expect in the later phase of the evolution of the Universe if
the density perturbations in the ideal fluid were produced in the very
early Universe.
  
Secondly, we will assume that the energy density perturbations in the
ideal fluid and the scalar field are related by the adiabatic
condition \cite{KoDa}.  Standard models of inflation \cite{LL} always
give this type of perturbation, as there is only one dynamical degree
of freedom during inflation, and so it is by far the most natural
choice to make. The entropy perturbation should therefore vanish,
\begin{equation}
\label{adiabcond1}
S_{\gamma\phi}\equiv\frac{\epsilon_{\gamma}}{\mu_{\gamma}+p_{\gamma}}-
	\frac{\epsilon_{\phi}}{\mu_{\phi}+p_{\phi}}=0\,,
\end{equation}
together with its first time derivative, 
\begin{equation}
\label{adiabcond2}
\dot{S}_{\gamma\phi}\equiv\frac{k^{2}}{a^{2}}\Psi_{\gamma\phi}-
	3He_{\gamma\phi}=0\,,
\end{equation}
where
\begin{equation}
\label{RelPsi}
\Psi_{\gamma\phi}\equiv\frac{\Psi_{\gamma}}{\mu_{\gamma}+p_{\gamma}}-
	\frac{\Psi_{\phi}}{\mu_{\phi}+p_{\phi}}\,,
\end{equation}
and
\begin{equation}
\label{Relent}
e_{\gamma\phi}\equiv\frac{e_{\gamma}}{\mu_{\gamma}+p_{\gamma}}-
	\frac{e_{\phi}}{\mu_{\phi}+p_{\phi}}\,.
\end{equation}
The quantities 
\begin{equation}
e_{\gamma} \equiv \varpi_{\gamma}-\frac{\dot{p}_{\gamma}}
	{\dot{\mu}_{\gamma}}\epsilon_{\gamma} \quad ; \quad
	e_{\phi} \equiv \varpi_{\phi}-\frac{\dot{p}_{\phi}}
	{\dot{\mu}_{\phi}}\epsilon_{\phi}\,, 
\end{equation}
represent the internal entropy of each component. In the case of the
ideal fluid we have $e_{\gamma}=0$.  {}From the adiabatic conditions
we obtain the initial values of $\delta{\phi}$ and $\delta{\phi}'$,
\begin{eqnarray}
\label{AdiabCond1}
\delta\phi&=&\frac{-2a\phi'V,_{\phi}\epsilon_{\gamma}+k^{2}H\phi'^{2}
	\Psi_{\gamma}} {(3\gamma\mu_{\gamma})[2V,_{\phi}-
	k^{2}\phi'/(3a)]}\,,\\
\label{AdiabCond2}
\delta\phi'&=&\phi'\frac{\epsilon_{\gamma}}{\gamma\mu_{\gamma}}-
	\frac{\delta\phi V,_{\phi}}{a^{2}H^{2}\phi'}+\phi'\alpha\,.
\end{eqnarray}

We also need to specify the initial values of the background variables
$H$, $\phi$ and $\phi'$ which determine the cosmological model.  The
initial value of $\mu_{\gamma}$ is obtained by the requirement that
the Universe has critical density. The other three degrees of freedom
for the initial background conditions are fixed by requiring specific
present values of the cosmological quantities $h$, $t_{0}$ and
$\Omega^{0}_{\gamma}$.  Note that for some $V(\phi)$, and given the
required values for $h$ and $\Omega^{0}_{\gamma}$, it may not be
possible to obtain the desired value for $t_{0}$. In general, for
fixed $h$ and $\Omega^{0}_{\gamma}$ there will be a maximum value
$t_{0}$ that can be reached for a given $V(\phi)$.

\section{The choice of scalar field potential}

Before we go on to perform the numerical integration of the
perturbation equations in both the ZSG and the UCG, we need to specify
what type of ideal fluid we will consider and the shape of the scalar
field potential. We are mainly interested in the growth of matter
density perturbations during the matter dominated era, and there are
good reasons to believe that most matter in the Universe has
negligible intrinsic velocity, i.e.~it is {\em cold}. In this paper we
will assume that the ideal fluid has no pressure, that is $\gamma=1$.

With relation to the type of scalar field we should consider the issue
is not as clear. Many different scalar fields have been proposed, most
of them with the specific aim of producing an inflationary expansion
phase in the very early Universe, though some also originate from
attempts at extending the Standard Model of particle physics.

We will consider two different scalar field potentials, an
exponential potential of the form
\begin{equation}
\label{exppot}
V(\phi)=V_{0}\exp(-\beta\phi)\,,
\end{equation}
and the potential associated with a pseudo-Nambu-Goldstone-boson
(PNGB) field \cite{HRoss},
\begin{equation}
\label{PNGBpot}
V(\phi)=M^{4}\left[\cos(\phi/f)+1\right]\,.
\end{equation}
Both these potentials have been extensively studied, within the context
of power-law inflation \cite{PowLaw} and natural
inflation \cite{NatInf} respectively. The PNGB field has also been proposed
explicitly as the most natural candidate for a presently-existing
minimally-coupled scalar field \cite{FCAI,CDFri}.

In both cases we have two degrees of freedom, and in principle one
should explore the full 2-dimensional parameter space defined by them.
However, our main objective in this paper is to draw attention to the
importance of taking into account the possibility of spatial
perturbations in a cosmological scalar field when one is assumed to
exist, due to their influence on the growth of matter density
perturbations. Therefore, in this paper we will only consider two sets
of values for the constants associated with each of the two
potentials.

In two of the models thus obtained, one for each potential, the scalar
field presently behaves like a slowly-decaying cosmological constant.
In these models, we will choose the initial values of the background
variables $H$, $\phi$ and $\phi'$, and the constants associated with
the potentials, so that we end up with $h = 0.6$, $\Omega^{0}_{\rm
m}=0.4$, and an age for the Universe of $t_{0}=14\;{\rm Gyrs}$. In the
other two models, again one for each potential, the scalar field
starts behaving like non-relativistic matter, scaling as $a^{-3}$, at
a redshift close to 100. Therefore, the age of the Universe in these
models is pretty much the same as if the Universe was always matter
dominated. In order to obtain an acceptable age we lower $h$ slightly
to $h = 0.55$ to give $t_{0}\simeq12\;{\rm Gyrs}$.

Our numerical integration of the perturbation equations will begin at
a redshift of $z=1100$, roughly at electron--photon decoupling, and end
at the present time. One reason for this choice is that through the COBE 
satellite 
measurement of the amplitude of
cosmic microwave background anisotropies we have good knowledge of the 
amplitude of energy density
perturbations existing at the horizon scale at this redshift
\cite{COBE}. The main reason though is that for reasonable parameter values 
we are well into matter domination and the effects of radiation on the 
matter power spectrum of density perturbations have already run their 
course. At this redshift, the power 
spectrum is well described by the cold dark matter transfer function, for 
example as parametrized by Bardeen et al.~\cite{BBKS,Sug}. We are thus able 
to analyze the effects of the scalar field on the matter power spectrum, 
without them being concealed within the full Boltzmann code machinery. We 
cannot however make predictions for the full 
microwave anisotropy power spectrum.

\subsection{The exponential potential}

For the exponential potential there is a particularly interesting
situation, which we will call EXP1, where the relative energy
densities of the scalar field and the ideal fluid remain constant with
time, thus implying $p_{\phi}=(\gamma-1)\mu_{\phi}$, after a
transitional period.  The energy density associated with the scalar
field is then a fixed fraction, $24\pi G\gamma/\beta^{2}$, of the
total energy density in the Universe \cite{Wett88,WCL,FerrJoy}. This
scaling solution of the cosmological background equations is one of
two attractor points for the system, the other being the well-known
power-law inflation solution. The former is the one which is attained
if the potential is steep enough. As homogeneous perturbations around
the scaling solution typically have complex eigenvalues, the system
usually approaches the attractor point through oscillations in the
relative energy densities of the scalar field and the ideal fluid. It
has recently been studied by Ferreira and Joyce \cite{FerrJoy}, though
for different parameters than the model we will look at.

That the Universe eventually reaches the scaling solution with
$\Omega_{\rm m}=0.4$ demands that $\beta=\sqrt{40\pi G}$. We choose the 
scalar field energy density at the start of the simulation to be much 
smaller than that of the 
scaling solution; this is not necessary, though it does have to be true much 
earlier at nucleosynthesis \cite{WCL,FerrJoy}.
This restricts the possible combinations for the values of
$\dot{\phi}$ at $z=1100$ and $V_{0}$. The value $\phi$ takes at
$z=1100$ is a matter of definition and we set it to zero. We will
assume $\dot{\phi}$ to be extremely small at $z=1100$, thus implying
that $V_{0}=(0.025\;{\rm eV})^{4}$. With this choice of parameters the
scalar field begins to contribute significantly to the total energy
density in the Universe by a redshift of about 70. In accordance with
the discussion above, this model therefore gives the same age as the
Einstein--de Sitter case, and to make it large enough we choose $h = 0.55$.

The second situation we consider, denoted EXP2, is where the system is
presently entering a period of power-law inflationary expansion,
analogous to the exponential expansion which would arise if a
cosmological constant is present in the Universe. To obtain our
desired values (stated above) for $\Omega^{0}_{\rm m}$, $h$ and
$t_{0}$ implies roughly that $\beta=\sqrt{28\pi G}$. On the other
hand, these constraints only very weakly define the initial splitting of the
energy density associated with the scalar field at $z=1100$ into its
potential and kinetic parts. We will again arbitrarily assume that
$\dot{\phi}(z=1100)\simeq 0$, thus implying $V_{0}=(0.0025\;{\rm
eV})^{4}$. 

\subsection{The PNGB potential}

In the case of the PNGB potential we will also examine two distinct
possible evolution histories for the Universe. In the model we will
call PNGB1, where $M=0.04\,h^{1/2}\;{\rm eV}$ and
$f=1.25\times10^{18}\;{\rm GeV}$, the scalar field starts to roll down
its potential at a redshift of about 100, and is presently coherently
oscillating at the bottom of the potential, behaving dynamically as
cold matter. This model, like EXP1, gives the same age as a Universe
containing only matter, so again we take $h$ to be 0.55, giving an age
for the Universe of 12 Gyrs. The constraint $\Omega^{0}_{\rm m}=0.4$
implies that $\phi(z=1100)=f$ and $\dot{\phi}(z=1100)\simeq0$.

In model PNGB2, where $M=0.003\,h^{1/2}\;{\rm eV}$ and $f=1/\sqrt{8\pi
G}=2.4\times10^{18}\;{\rm GeV}$, the scalar field begins moving down
its potential at around $z=1$, and has not yet reached the bottom of
the potential by the present time. The values of $M$ and $f$ in this
model coincide with those chosen for a more detailed analysis by
Frieman et al.~\cite{FCAI} in their paper dealing with PNGB motivated
dynamical cosmological constant models.\footnote{In a more recent
paper, Coble et al.~\cite{CDFri} considered a slightly different model
with $M=0.005\;{\rm eV}$ and $f=1.885\times10^{18}\;{\rm GeV}$, where
the scalar field starts moving down the PNGB potential at $z\simeq10$
and is presently already oscillating around its minimum. These values
were chosen so that the model yields $\Omega^{0}_{\rm m}=0.4$ and
$h=0.7$.  This implies an age for the Universe only slightly above 10
Gyr.  We prefer the PNGB2 model over theirs. If one chose to obtain
$h=0.55$, and thus an age close to 12 Gyr, one would end up with
$\Omega^{0}_{\rm m}=0.5$.} Imposing the boundary condition
$\Omega^{0}_{\rm m}=0.4$ leads to $\phi(z=1100)=1.75f$ and
$\dot{\phi}(z=1100)\simeq0$, while $h=0.6$ and $t_{0}=14\;{\rm Gyrs}$
are obtained by simply choosing the correct value for $H$ at $z=1100$.

\section{Results}

We will work in the 
ZSG, and display the fluid density perturbation 
$\delta_{\gamma}^{\chi}\equiv\epsilon_{\gamma}^{\chi}/\mu_{\gamma}$. We 
arbitrarily take its initial value at redshift 1100 to be $10^{-5}$;
as the equations are linear, and our main results only show power spectra 
relative to one another, the initial  
magnitude of the density perturbation is irrelevant.  Note 
that the density perturbations given in different gauges coincide on scales 
well within the horizon, but otherwise, though uniquely defined as long as 
the gauge is given, do not coincide. Care must therefore be taken in 
interpreting any long-wavelength behaviour. Most of the literature uses the 
comoving gauge if the large-scale power spectrum is shown.

We integrated the background and
perturbation equations both in the ZSG and the UCG using the NAG
Fortran Library Routine D02CJF, which is based on a variable-order,
variable-step Adams method, for values of $k$ in the range of
$10^{-5}$ to $1\, h\,{\rm Mpc}^{-1}$. We did this not only for the four 
scalar field models chosen in the previous section, but also 
for two cases where no scalar field was assumed present, which will 
serve as comparison. In one the Universe has critical density, while in the 
other the Universe is flat with $\Omega_{0}=0.4$, thus implying the 
presence of a cosmological constant.

\begin{figure*}[t]
\begin{center}
\leavevmode\epsfysize=9.5cm \epsfbox{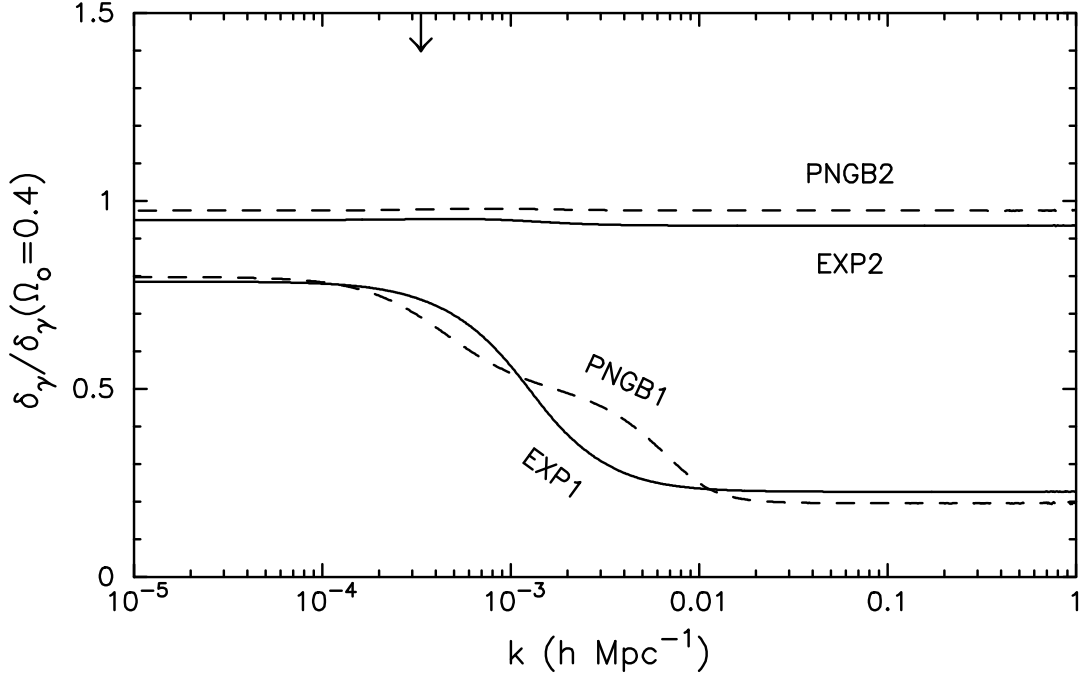}\\ 
\end{center}
\caption[Relative perturbation growth]{\label{Fig1} This shows the present 
amplitude of fractional perturbations in the energy density of the
ideal fluid in the ZSG for our four different models, relative to the
one obtained in the case where no scalar field is present and the
Universe is flat with $\Omega_{0}=0.4$. The full and dashed lines represent 
respectively the two exponential potential and the two PNGB models. 
The arrow indicates the inverse Hubble radius, $k=aH$, at present.}
\end{figure*} 

In Fig.~1 we plot the present amplitude obtained for 
$\delta_{\gamma}^{\chi}(k)$ (for simplicity henceforth omitting the 
superscript and always meaning this gauge) for our four models, relative to
that found in the case of a flat Universe with $\Omega_{0}=0.4$ 
(with $h$ in this comparison model adjusted to match that of each of the 
four scalar field models we consider).   
The main integration runs were done in the ZSG, but when the same runs 
were performed in the UCG, with $\epsilon_{\gamma}^{\varphi}(z=1100)$ 
calculated from $\epsilon_{\gamma}^{\chi}(z=1100)$ using 
Eq.~(\ref{grelUCG3}) and then $\epsilon_{\gamma}^{\chi}(z=0)$ from
$\epsilon_{\gamma}^{\varphi}(z=0)$ using Eq.~(\ref{grelZSG3}),
the difference in the present value of $\epsilon_{\gamma}^{\chi}$ was
less than one per cent over the whole range for $k$ in the four models. 

Although we believe that Eqs.~(\ref{AdiabCond1}) and 
(\ref{AdiabCond2}) should be used to obtain the initial values for 
$\delta\phi$ and $\delta\phi'$, in accordance with the adiabatic 
relations Eqs.~(\ref{adiabcond1}) and (\ref{adiabcond2}) one expects 
from inflationary generated perturbations, we also looked at what 
occurs when arbitrary initial values for these quantities are assumed. 
After performing several integration runs for a variety of initial values 
for $\delta\phi$ and $\delta\phi'$, we reached the conclusion that 
the evolution of the energy density perturbations in the 
ideal fluid is almost (to a few per cent) independent of the initial 
values one considers for these quantities, 
as long as they are less than of order unity. In particular, this 
holds true (to better than 1 per cent) when it is assumed that 
there are no initial scalar field perturbations either in the ZSG, i.e. 
$(\delta\phi')^{\chi}=\delta\phi^{\chi}=0$, or in the UCG, i.e. 
$(\delta\phi')^{\varphi}=\delta\phi^{\varphi}=0$. So the initial condition 
for the scalar field perturbations does not seem particularly important.

In Fig.~2 we plot the evolution of $\delta_{\gamma}(a)$ for 
perturbations with a comoving wavenumber of 
$k=0.5\, h \,{\rm Mpc}^{-1}$ between $z=100$ and the present. 
All the four scalar field models are shown, together with the cases 
of a critical-density Universe and a flat Universe with 
$\Omega_{0}=0.4$. The integration runs were actually started at 
$z=1100$, but 
all models behave very much as the critical-density case 
up to $z=100$. We chose the wavenumber shown as it corresponds 
to a scale which was already well inside the Hubble radius at the 
beginning of the integrations, so we need not worry about specifying the 
gauge. In fact, 
comparing the present amplitude of $\delta_{\gamma}$ 
obtained in the ZSG and in the UCG for the various models 
under consideration, we find that the density perturbations in the two 
gauges coincide for values of $k$ down to 
about $10^{-3}\, h\,{\rm Mpc}^{-1}$, and can therefore be regarded as 
gauge-independent. For smaller $k$ values one needs to be careful to specify 
the gauge used.

\begin{figure*}[t]
\begin{center}
\leavevmode\epsfysize=9.5cm \epsfbox{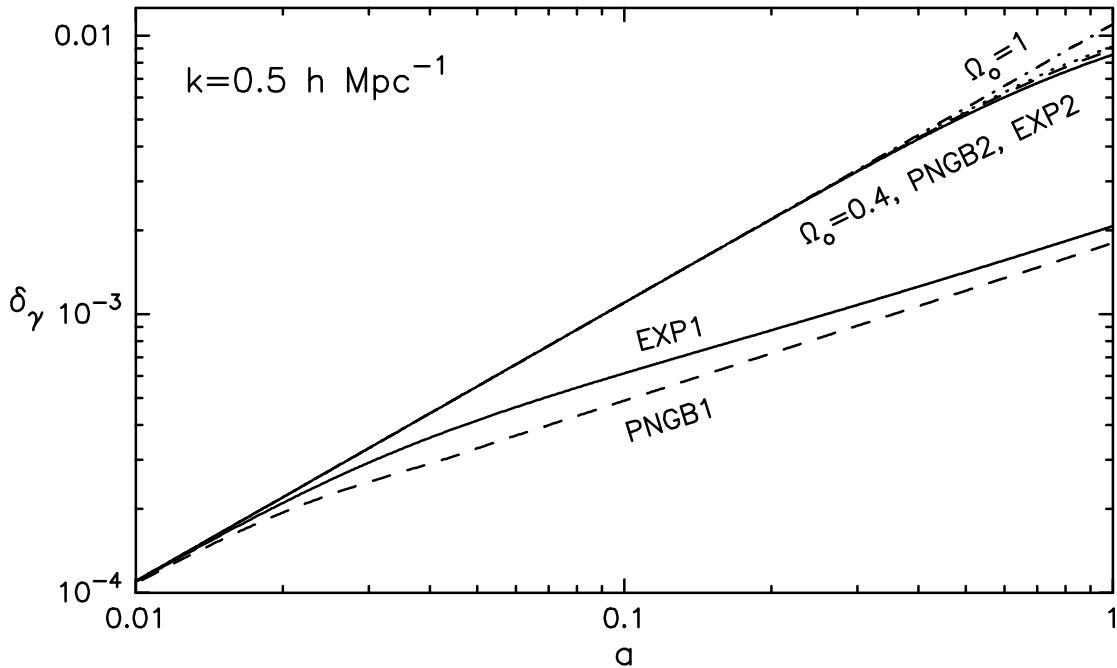}\\ 
\end{center}
\caption[Absolute perturbation growth]{\label{Fig2} The 
evolution in the ZSG of the perturbations with a comoving wavenumber of 
$k=0.5\, h\,{\rm Mpc}^{-1}$. The dash-dotted and dotted lines represent the 
cases where respectively the Universe has critical density and the Universe 
is flat with $\Omega_{0}=0.4$, with no scalar field present. Solid lines 
show the EXP models and dashed ones the PNGB models.}
\end{figure*} 

As mentioned in Section III, Fig.~1 represents the distortion in the square 
root of the matter power spectrum suffered 
since redshift $1100$, $P(k)\propto \delta_{\gamma}^{2}$, relative to a flat 
Universe with $\Omega_{0}=0.4$ and no scalar field.
Since the full spectrum in that model, including all known effects of 
radiation and neutrinos, is well known, we can therefore use it to obtain 
the expected present shape of $P(k)$ for our four scalar field models. 

We show the power spectra in Fig.~3, where we simply multiplied the expected 
present shape of $P(k)$ in a flat Universe with $\Omega_{0}=0.4$, 
$\Omega_{\rm B}=0.016\,h^{-2}$ and either $h=0.55$ or $h=0.6$, with no 
scalar field present, by 
$\delta^2_{\gamma}/\delta^2_{\gamma}(\Omega_{0}=0.4)$. This procedure is 
correct since up to $z=1100$ the background and perturbation 
evolution of any of the scalar field models is equivalent to that of 
a cosmological constant model which has the same values for 
$\Omega_{0}$ and $h$ as each individual scalar field model. We also show 
in Fig.~3 the expected present shape of $P(k)$ in a critical-density 
Universe for the two values of $h$ considered. 
We obtained the expected present shape of $P(k)$ for both the 
critical density and the cosmological constant cases 
by assuming a Harrison--Zel'dovich (scale-invariant) 
primordial power spectrum for the energy density perturbations and 
using the cold dark matter transfer function initially derived by 
Bardeen et al. \cite{BBKS}, and later modified by Sugiyama \cite{Sug} to 
include the contribution of baryons. 

The critical-density and cosmological constant models are normalized to COBE 
\cite{COBE}. The amplitude of the other models, which has been computed 
relative to the latter, is almost correct. It includes the two main effects 
--- that the early time matter power spectrum in low-density models is 
higher by a factor $1/\Omega_0$ than in critical-density models is encoded 
in the initial conditions through the different initial value of $H$, and 
the growth suppression factor from the scale factor dynamics is computed in 
the subsequent evolution. As each model (except the critical-density one) 
has the same redshift of matter--radiation equality, all that is omitted is 
the line-of-sight contribution to the Sachs--Wolfe effect, which requires a 
full Boltzmann code for accurate computation. The effect of this term on the 
COBE normalization is non-existent for the critical-density case, and known 
to be negligible in the cosmological constant case (see 
e.g.~Ref.~\cite{LLreion,LLVW}). It may be slightly more significant for the 
decaying cosmological constant case \cite{CDFri} and future observations 
would probably require an accurate computation, while present ones do not.

We only show the power spectra for values of $k$ larger than 
$10^{-3}\, h\,{\rm Mpc}^{-1}$. On smaller scales the amplitude of the 
power spectrum is gauge-dependent, though of course well-defined in any 
particular gauge. 

\begin{figure*}[t]
\begin{center}
\leavevmode\epsfysize=5.5cm \epsfbox{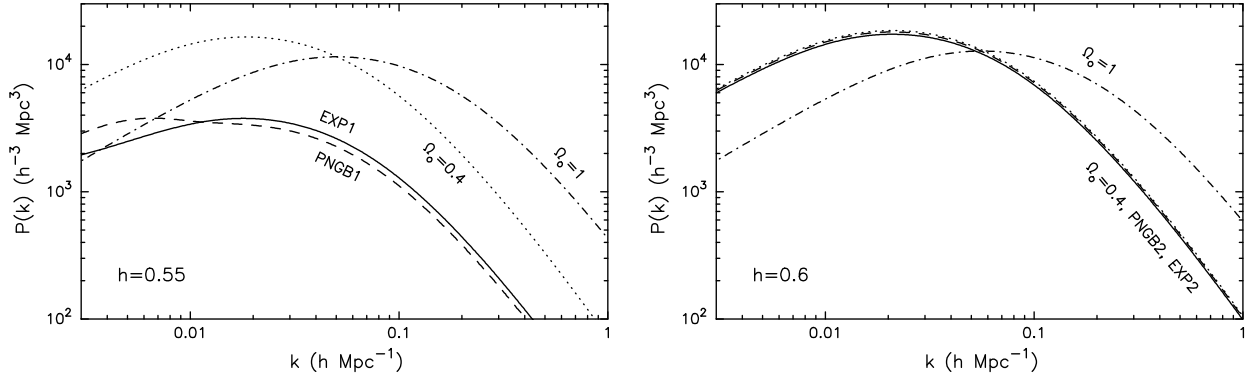}\\ 
\end{center}
\caption[Power Spectrum]{\label{Fig3} The power spectrum of the energy 
density perturbations in the matter component for the models, all 
normalized to COBE as described in the text. The line  
styles represent $P(k)$ for the same situations as in the 
previous two figures. Note that the PNGB2 and EXP2 models give a very 
similar spectrum to the standard cosmological constant case.}
\end{figure*}

\section{Discussion}

The results presented in this paper extend previous work on the 
effects of a presently-existing scalar field, with regard to 
the evolution of energy density perturbations in a 
pressureless fluid. In accordance with general covariance we 
allow for the presence of spatial perturbations in the scalar field. 
Contrary to previous authors, we explicitly relate the energy density 
perturbations in the pressureless fluid to those in the scalar field 
through the adiabatic condition, as expected if those perturbations 
arose from a period of inflation in the very early Universe. 

The study of the perturbation evolution was performed both in the zero-shear 
and uniform-curvature gauges, in the latter case for the first time. 

We considered two possible potentials for the scalar field, an 
exponential and that associated with a 
pseudo-Nambu-Goldstone-boson field. Each of these 
potentials has two degrees of freedom, and we worked with two 
different sets of parameters for each potential. Those models, 
EXP2 and PNGB2, which were chosen so that 
the scalar field becomes dynamically important only very recently, 
at $z\sim1$, and is presently behaving as a cosmological constant, 
yield a shape for the power spectrum of energy density 
perturbations in the matter component extremely close to that 
one obtains in a cosmological constant model where 
$\Omega_{0}=0.4$ and $h=0.6$, though the amplitude 
is a few percent smaller. 

The situation is rather different for the other two models, 
EXP1 and PNGB1, chosen so 
that the scalar field becomes dynamically important at a 
much earlier time, around a redshift of 100. They presently 
behave as pressureless matter. In the case of model EXP1 
the shape of the power spectrum of energy density 
perturbations in the matter component is again very close to that 
one obtains in a cosmological constant model, this time with 
$\Omega_{0}=0.4$ and $h=0.55$. Its amplitude is however 
only about 5 per cent of that for the cosmological constant 
model. Model PNGB1 has the further interesting feature that the shape of 
$P(k)$ 
is altered for values of $k$ below $0.01\, h\,{\rm Mpc}^{-1}$. 
Unfortunately this corresponds to scales above around  
$100\, h^{-1}\,{\rm Mpc}$, 
at the limit of what present cluster and galaxy surveys 
can probe. The feature in the shape of $P(k)$ for model PNGB1 is 
due to oscillations in the rate of decrease of the Hubble parameter when 
the scalar field becomes dynamically important at $z\sim100$. 
On smaller scales the suppression of the growth of the 
energy density perturbations in the matter component is 
similar to that in model EXP1.

The results just described suggest that the presence of 
spatial perturbations in a presently-existing scalar 
field substantially affects the evolution of 
energy density perturbations in the matter component only if  
the scalar field has contributed significantly to the total energy 
density in the Universe for several Hubble times. If the 
scalar field is becoming dynamically important only now, the presence of 
spatial perturbations does not seem to have much effect. 

The similarity between the scalar field models EXP2 and PNGB2, 
and a flat model with $\Omega_{0}=0.4$, $h=0.6$ and no scalar field, makes 
the first two models attractive substitutes of the latter. Further, given 
that 
the above mentioned flat model is able to reproduce all the most 
reliable observational data presently available on large-scale structure 
and cosmic microwave background temperature anisotropies \cite{LLVW}, 
we expect models EXP2 and PNGB2 to do as well. As they are quite similar to 
the standard cosmological constant case, a detailed evaluation of the 
supernova constraint would be interesting; Frieman et al.~\cite{FCAI} 
suggest that the constraint will weaken so they should be viable.

With regard to models EXP1 and PNGB1, their much smaller growth for the 
matter perturbations seems very problematic. We expect the dispersion of the 
density contrast smoothed on spheres of $8\, h^{-1}\,{\rm Mpc}$, usually 
represented by $\sigma_{8}$, required for both models so that they can 
reproduce the present abundance of high mass X-ray emitting galaxy clusters 
to be the same as for the critical-density case, due to the two models being 
dynamically equivalent to it since well before redshift 10. This 
conservatively requires 
$\sigma_{8}$ in the range 0.45 to 0.8 for both models \cite{galclu}. On the 
other hand, as 
the growth suppression factor of the matter perturbations is about 4.5 times 
larger in the EXP1 and PNGB1 models than in a flat model with 
$\Omega_{0}=0.4$ and $h=0.55$, to a first approximation this means that the 
value of $\sigma_{8}$ implied by COBE for the two scalar field models would 
be about 4.5 times smaller than for the flat model \cite{CDFri}.  
Inclusion of the integrated Sachs--Wolfe effect may lead to a slight further 
decrease in $\sigma_{8}$ \cite{CDFri}. If the primordial power spectrum of 
energy density 
perturbations is assumed scale-invariant, then this would mean that 
$\sigma_{8}<0.18$ for both EXP1 and PNGB1 \cite{LLVW}. 
One would need a very `blue' primordial power spectrum, with at least 
$n>1.45$, for $\sigma_{8}$ to reach the minimum requirement of 0.45. 
Ferreira and Joyce \cite{FerrJoy} advocate a much larger value of 
$\Omega_0$, which resolves the amplitude problem but raises the question of 
whether the change in the shape of the spectrum can really be enough. We are 
presently carrying out a full investigation of all these models against a 
range of observational data.

As in both EXP1 and PNGB1 the scalar field goes through a period at a 
redshift of about 100 when it behaves dynamically like a cosmological 
constant, we expect the full angular anisotropy spectrum for the cosmic 
microwave background radiation in these models to display a  
distinctive signature. This clearly merits detailed investigation.

\vspace*{8pt}
{\em Note:} As we were completing this paper, a preprint by Caldwell
et al.~\cite{CDS} appeared which covers similar issues. They use an enhanced 
Boltzmann code to generate microwave anisotropy power spectra. Their matter 
power spectra appear in good agreement with ours. They did not use adiabatic 
initial conditions for the scalar field perturbations, but we have shown 
that this is unlikely to have significant effects.

\acknowledgments

P.T.P.V.~was supported by PPARC and A.R.L.~by the Royal Society. We
thank Scott Dodelson, Pedro Ferreira and Martin White for discussions, and 
Wo-Lung Lee for pointing out an error in an earlier version of this paper.
We acknowledge use of the Starlink computer system at the University of
Sussex.
\appendix

\section{The evolution equations}

\subsection{Gauge considerations}

A cosmological perturbation is defined by means of a correspondence
between an arbitrary background spacetime and the real physical
Universe.  A {\em gauge transformation} is a change in this
correspondence, keeping the background spacetime fixed. Therefore, in
general the value of the perturbed part of some physical quantity will
not be invariant under a gauge transformation. Further, the degrees of
freedom due to gauge transformations give rise to spurious unphysical
modes in the solutions to the evolution of the perturbed part of
gauge-dependent quantities, which can always be removed by a
convenient gauge transformation. In order to avoid these spurious
modes one either evolves gauge-invariant quantities related to the
gauge-dependent quantities one is actually interested in, obtaining
the latter from the former at any one time, or one has to provide a
gauge-fixing condition which completely specifies the way through
which spacetime is to be split into background and perturbed
components.

The linear analysis of cosmological perturbations was initiated by
Lifshitz, with a seminal paper published in 1946 \cite{Lif}, who used
the so-called synchronous gauge-fixing condition. Unfortunately this
condition, though considerably simplifying the perturbation equations,
still leaves a residual gauge degree of freedom. The spurious modes
thus arising are difficult to distinguish from the real physical ones,
and their identification was a source of controversy for some time.
The use of gauge-invariant quantities in the calculation of the
evolution of cosmological perturbations only really took off with the
paper by Bardeen in 1980 \cite{Bar}.  Though it avoids the problem of
obtaining unphysical modes in the solutions to the perturbation
equations, it really does not offer any extra advantage over the
gauge-specific methods which remove any gauge degree of freedom by
completely fixing the background/perturbed splitting.

In this paper we will use gauge-specific methods to solve the
perturbation equations. We will adopt the notation and use the
equations laid down in a series of papers by Hwang
\cite{Hw1,Hw2,Hw3,Hw4,Hwetal}. We will be solely interested in the
evolution of density (scalar) perturbations. The system is composed of
an ideal fluid plus a single minimally coupled real scalar field,
$\phi$, evolving in a background Einstein-de Sitter Universe.  Given
that the spatial part of the background spacetime is thus homogeneous
and isotropic, the perturbations in any physical quantities will
necessarily be gauge-invariant under purely spatial gauge
transformations. We will therefore only worry about the temporal gauge
transformation.

\subsection{Notation and general equations}

We will now introduce the gauge non-specific perturbation equations
obtained by Hwang. They relate the perturbed part of the metric
variables, $\alpha$ (perturbed part of the lapse function), $\varphi$
(perturbed part of the spatial curvature), $\chi$ (perturbed part of
the shear) and $\kappa$ (perturbed part of the expansion scalar), to
the perturbed part of the matter variables,
$\epsilon=\epsilon_{\gamma}+\epsilon_{\phi}$ (perturbed part of the
total energy density), $\varpi=\varpi_{\gamma}+\varpi_{\phi}$
(perturbed part of the total pressure)\footnote{Here we changed the
notation from $\pi$ to $\varpi$ to avoid confusion with the number
$\pi$.} and $\Psi=\Psi_{\gamma}+\Psi_{\phi}$ (perturbed part of the
total energy density flux, or total fluid four-velocity, depending on
the frame chosen). We have
\begin{equation}
\varpi_{\gamma}=(\gamma-1)\epsilon_{\gamma}\,,
\end{equation}
and one can derive that \cite{Hw4}
\begin{eqnarray}
\label{pertener}
\epsilon_{\phi}&=&\dot{\phi}\dot{\delta\phi}-\dot{\phi}^{2}\alpha+
	V,_{\phi}\delta\phi\,,\\
\label{pertpress}
\varpi_{\phi}&=&\dot{\phi}\dot{\delta\phi}-\dot{\phi}^{2}\alpha-
	V,_{\phi}\delta\phi\,,\\
\label{pertvel}
\Psi_{\phi}&=&-\dot{\phi}\delta\phi\,,
\end{eqnarray}
where $\delta\phi$ is the perturbed part of the scalar field.  We
will be particularly interested in the evolution of the fractional
perturbation in the energy density of the ideal fluid,
$\delta_{\gamma}\equiv\epsilon_{\gamma}/\mu_{\gamma}$.

We will express the perturbed parts of both the metric and matter variables 
by means of Fourier expansions. For example,
\begin{equation}
\delta\phi({\bf\rm x},t)=\sum_{k}\delta\phi_{\bf\rm k}(t)
	e^{i{\bf\rm k.x}}\,,
\end{equation}
where 
\begin{equation}
\delta\phi_{\bf\rm k}(t)=\frac{1}{V}\int\delta\phi({\bf\rm x},t)
	e^{-i{\bf\rm k.x}}d{\bf\rm x}\,,
\end{equation}
being $k\equiv|{\bf\rm k}|$ a fixed comoving wavenumber. The Fourier
expansions are made in a large enough box that the induced periodicity
is irrelevant.

As in this paper we are only interested in the linear evolution of
cosmological perturbations, we will assume that the different Fourier
modes for each variable behave independently of each other. We will
drop the suffices ${\bf\rm k}$ identifying each Fourier mode in order
to lighten the notation.

In the case of our system, formed by an ideal fluid plus a minimally
coupled real scalar field, the perturbation equations take the form
\cite{Hw1,Hw2,Hw4},
\begin{eqnarray}
\label{pert1}
3\dot{\varphi}& =& 3H\alpha-\kappa+\frac{k^{2}}{a^{2}}\chi\,,\\
\label{pert2}
-\frac{k^{2}}{a^{2}}\varphi+H\kappa & = & -4\pi G(\epsilon_{\gamma}+
	\dot{\phi}\dot{\delta\phi}-\dot{\phi}^{2}\alpha+
	V,_{\phi}\delta\phi)\,,\\
\label{pert3}
\kappa-\frac{k^{2}}{a^{2}}\chi & = & -12\pi 
	G(\Psi_{\gamma}-\dot{\phi}\,\delta\phi)\,,\\
\label{pert4}
\dot{\chi}+H\chi & = &\alpha+\varphi\,,\\
\label{pert5}
\dot{\kappa}+2H\kappa & = & \left(\frac{k^{2}}{a^{2}}-3\dot{H}\right)\alpha+
	4\pi G[(3\gamma-2)\epsilon_{\gamma}+
	4\dot{\phi}\dot{\delta\phi}-
	4\dot{\phi}^{2}\alpha-2V,_{\phi}\delta\phi]\,,\\
\label{pert6}
\dot{\epsilon_{\gamma}}+3H\gamma\epsilon_{\gamma} & = &
	\gamma\mu_{\gamma}(\kappa-3H\alpha)+
	\frac{k^{2}}{a^{2}}\Psi_{\gamma}\,,\\
\label{pert7}
\dot{\Psi}_{\gamma}+3H\Psi_{\gamma} & = & -\gamma\mu_{\gamma}\alpha-
	(\gamma-1)\epsilon_{\gamma}\,,\\
\label{pert8}
\ddot{\delta\phi}+3H\dot{\delta\phi}+\left(\frac{k^{2}}{a^{2}}+V,_{\phi\phi}
	\right)
\delta\phi & = & \dot{\phi}(\kappa+\dot{\alpha})
	-(3H\dot{\phi}+2V,_{\phi})\alpha\,.
\end{eqnarray}
It should again be stressed that these equations were obtained without
reference to any gauge-fixing condition.  No anisotropic pressure term
appears as in both the case of an ideal fluid and a minimally coupled
real scalar field the anisotropic pressure is zero. The last three
equations are, in order, the energy and momentum conservation
equations for the perturbations in the ideal fluid, and the energy
conservation equation for the perturbations in the scalar field.  The
momentum conservation equation for the perturbations in the scalar
field is identically satisfied.

The most obvious and fundamental gauge-fixing conditions follow from
requiring that the perturbed part of one of the metric or matter
variables is zero.  We thus have: the synchronous gauge,
$\alpha\equiv0$; the uniform-curvature gauge, $\varphi\equiv0$; the
zero-shear gauge, $\chi\equiv0$; the uniform-expansion gauge,
$\kappa\equiv0$; the uniform-density gauge, $\epsilon\equiv0$; the
uniform-pressure gauge, $\varpi\equiv0$; and the comoving gauge,
$\Psi\equiv0$. Except for the synchronous gauge, all the other
gauge-fixing conditions completely remove the gauge modes from the
solutions to the perturbation equations. We will use two of these
gauge-fixing conditions to derive two different sets of perturbation
equations from the system given above. With the aid of expressions
relating quantities in the two gauges we will thus be able to estimate
the numerical errors arising from the numerical integration of both
sets of perturbation equations.  We will consider the zero-shear gauge
(ZSG, also known as the longitudinal or conformal Newtonian gauge
\cite{MukFB}), and the uniform-curvature gauge (UCG). These choices
are the ones which lead to the two simplest sets of perturbation
equations, thus decreasing the probability of numerical errors
creeping into the solutions. The two sets can be obtained by simply
getting rid of $\chi$ and $\dot{\chi}$ in the case of the ZSG, and
$\varphi$ and $\dot{\varphi}$ for the UCG, in Eqs.~(\ref{pert1}) to
(\ref{pert8}).

\section{Numerical solutions}

\subsection{Background and perturbation equations}

The evolution of the background variables $H$ and $\phi$ is obtained
by numerically solving the system of first-order differential
equations formed by Eq.~(\ref{Friedmann}) and the two first-order
differential equations that can be obtained from Eq.~(\ref{enconssf}),
\begin{eqnarray}
\label{back1}
\frac{d \phi}{d a}&=&f\,,\\
\label{back2}
\frac{d f}{d a}&=&-4\frac{f}{a}+
	4\pi G\left(\frac{\gamma\mu_{\gamma}f}{aH^{2}}+af^{3}\right)-
	\frac{V,_{\phi}}{a^{2}H^{2}}\,,\\
\label{back3}
\frac{d H}{d a}&=&-4\pi G\left(aHf^{2}+
	\frac{\gamma}{aH}\mu_{\gamma}\right)\,,
\end{eqnarray}
where $\mu_{\gamma}$ is given by analytically solving Eq.~(\ref{enconsfl}),
\begin{equation}
\label{energyevol}
\mu_{\gamma}=\mu_{\gamma}^{0}\left(\frac{a}{a_{0}}\right)^{-3\gamma}\,.
\end{equation}
The suffix `0' indicates present-day values as usual.  Note that the
independent variable has been changed from coordinate time, $t$, to
the scale factor, $a$. They are related by the first-order
differential equation
\begin{equation}
\label{age}
\frac{d t}{d a}=a^{-1}H^{-1}\,,
\end{equation}
the integration of which gives the time elapsed in the Universe
between two given values of the scale factor. Derivatives with respect
to the scale factor will be represented by a prime. We choose the
independent variable to be the scale factor as it is easier to work
with numerically and is more meaningful from the point of view of
structure formation.

We have 8 perturbation equations for 11 dependent perturbation
variables in both the ZSG and the UCG: $\epsilon_{\gamma}$,
$\epsilon_{\gamma}'$, $\Psi_{\gamma}$, $\Psi_{\gamma}'$,
$\delta{\phi}$, $\delta\phi'$, $\delta{\phi}''$, $\alpha$ and $\kappa$
in either, along with $\varphi$ and $\varphi'$ in the ZSG, or $\chi$
and $\chi'$ in the UCG. Eqs.~(\ref{pert1}) to (\ref{pert4}),
(\ref{pert6}) and (\ref{pert7}) will be used to describe the evolution
of $\alpha$, $\kappa$ and the quantities associated with the ideal
fluid, $\epsilon_{\gamma}$, $\epsilon_{\gamma}'$, $\Psi_{\gamma}$ and
$\Psi_{\gamma}'$, in terms of $\delta\phi$, $\delta{\phi}'$, $\varphi$
and $\varphi'$ (ZSG) (the last two variables are replaced by $\chi$
and $\chi'$ in the UCG), and the background variables. We thus have in
the ZSG,
\begin{eqnarray}
\label{pertZSG1}
\alpha & = & -\varphi\,,\\
\label{pertZSG2}
\kappa & = & -3H\varphi-3aH\varphi'\,,\\
\label{pertZSG3}
\epsilon_{\gamma} & = & \frac{\varphi}{4\pi G}\frac{k^{2}}{a^{2}}+
	\frac{3H^{2}\varphi+3aH^{2}\varphi'}{4\pi G}-
	a^{2}H^{2}\phi'\delta\phi'-
	a^{2}H^{2}\phi'^{2}\varphi-V,_{\phi}\delta\phi\,,\\
\label{pertZSG4}
\Psi_{\gamma} & = & \frac{H\varphi+aH\varphi'}{4\pi 
	G}+aH\phi'\delta\phi\,,\\
\label{pertZSG5}
\epsilon'_{\gamma} & = & -3\gamma\frac{\epsilon_{\gamma}}{a}-
	3\gamma\mu_{\gamma}\varphi'+
	\frac{k^{2}}{a^{2}}\frac{\Psi_{\gamma}}{aH}\,,\\
\label{pertZSG6}
\Psi'_{\gamma} & = & -3\frac{\Psi_{\gamma}}{a}+
	\gamma\mu_{\gamma}\frac{\varphi}{aH}-
	(\gamma-1)\frac{\epsilon_{\gamma}}{aH}\,,
\end{eqnarray}
and for the UCG,
\begin{eqnarray}
\label{pertUCG1}
\alpha & = & H\chi+aH\chi'\,,\\
\label{pertUCG2}
\kappa & = & \left(3H^{2}+\frac{k^{2}}{a^{2}}\right)\chi+3aH^{2}\chi'\,,\\
\label{pertUCG3}
\epsilon_{\gamma} & = & 
	-\left(3H^{2}+\frac{k^{2}}{a^{2}}\right)\frac{H\chi}{4\pi G}-
	\frac{3aH^{3}\chi'}{4\pi G}-
	a^{2}H^{2}\phi'\delta\phi'+
	a^{2}H^{2}\phi'^{2}(aH\chi'+H\chi)-V,_{\phi}\delta\phi\,,\\
\label{pertUCG4}
\Psi_{\gamma} & = & -\frac{H^{2}\chi+aH^{2}\chi'}{4\pi 
	G}+aH\phi'\delta\phi\,,\\
\label{pertUCG5}
\epsilon'_{\gamma} & = & -3\gamma\frac{\epsilon_{\gamma}}{a}+
	\gamma\mu_{\gamma}\frac{k^{2}}{a^{2}}\frac{\chi}{aH}+
	\frac{k^{2}}{a^{2}}\frac{\Psi_{\gamma}}{aH}\,,\\
\label{pertUCG6}
\Psi'_{\gamma} & = & -3\frac{\Psi_{\gamma}}{a}-
	\gamma\mu_{\gamma}\frac{\chi+a\chi'}{a}-
	(\gamma-1)\frac{\epsilon_{\gamma}}{aH}\,.
\end{eqnarray}
Using these expressions we can now convert Eqs.~(\ref{pert5}) 
and (\ref{pert8}) into the following second-order differential equations, 
\begin{eqnarray}
\label{pertZSG7}
\delta\phi'' & = & 
	-\left(\frac{4}{a}+\frac{H'}{H}\right)\delta\phi'-
	\left(\frac{k^{2}}{a^{2}}+V,_{\phi\phi}\right)
	\frac{\delta\phi}{a^{2}H^{2}}-
	4\phi'\varphi'+2\frac{\varphi V,_{\phi}}{a^{2}H^{2}}\,,\\
\label{pertZSG8}
\varphi'' & 
	= & -\left[\frac{(3\gamma+2)}{a}+\frac{H'}{H}\right]\varphi'-
	\left[\frac{3\gamma}{a^{2}}+\frac{2H'}{aH}+
	(\gamma-1)\frac{k^{2}}{a^{4}H^{2}}-4\pi G(\gamma-2)\phi'^{2}
	\right]\varphi+\nonumber \\
 & & 4\pi G\left[(\gamma-2)\phi'\delta\phi'+
	\gamma \frac{\delta\phi V,_{\phi}}{a^{2}H^{2}}\right]\,,
\end{eqnarray}
in the ZSG and 
\begin{eqnarray}
\label{pertUCG7}
\delta\phi'' & = & -\left[\frac{4}{a}+\frac{H'}{H}+4\pi 
	G(\gamma-2)a\phi'^{2}\right] \delta\phi'-
	\left( \frac{k^{2}}{a^{2}}+V,_{\phi\phi}+
	4\pi G\gamma a\phi' V,_{\phi}\right)
	\frac{\delta\phi}{a^{2}H^{2}}\nonumber \\
 & & -\left[3\gamma H\phi'+2aH'\phi'+\frac{2V,_{\phi}}{aH}-
	4\pi G(\gamma-2)a^{2}H\phi'^{3}\right]\chi' \nonumber \\
 & & - \left[2H'\phi'+\frac{3\gamma 
	H\phi'}{a}+\frac{2V,_{\phi}}{a^{2}H}+
	(\gamma-2)\frac{k^{2}}{a^{2}}\frac{\phi'}{aH}-
	4\pi G(\gamma-2)aH\phi'^{3}\right]\chi\,,\\
\label{pertUCG8}
\chi'' & = & -\left[\frac{3\gamma+2}{a}+3\frac{H'}{H}-
	4\pi G (\gamma-2)a\phi'^{2}\right]\chi'-
	\left[\frac{3H'}{aH}+\frac{3\gamma}{a^{2}}+
	(\gamma-1)\frac{k^{2}}{a^{4}H^{2}}-
	4\pi G (\gamma-2)\phi'^{2}\right]\chi \nonumber \\
 & & -4\pi G\left[(\gamma-2)\frac{\phi'\delta\phi'}{H}+
	\gamma\frac{\delta\phi V,_{\phi}}{a^{2}H^{3}}\right]\,,
\end{eqnarray}
in the UCG. Each of these equations can be split into two
first-order differential equations, in the same way as we did for
Eq.~(\ref{enconssf}), which we will then numerically integrate in
order to determine the evolution of $\delta\phi$, $\delta{\phi}'$,
$\varphi$ and $\varphi'$ (ZSG) ($\chi$ and $\chi'$ in the UCG).

In all we will need to simultaneously numerically integrate a system
of seven first-order ordinary differential equations, formed by the
background Eqs.~(\ref{back1}) to (\ref{back3}), where expression
(\ref{energyevol}) gives $\mu_{\gamma}$, and the four perturbation
equations resulting from either (\ref{pertZSG7}) and (\ref{pertZSG8})
(in the ZSG), or (\ref{pertUCG7}) and (\ref{pertUCG8}) (in the UCG).
The question of initial conditions for this procedure is addressed in
the main text of this paper.

\subsection{Relations between quantities in different gauges}

Once we have calculated the evolution of the perturbation variables in
some particular gauge, we can use gauge-invariant variables and the
gauge non-specific set of perturbation Eqs.~(\ref{pert1}) to
(\ref{pert8}) to obtain the evolution of such variables in any other
gauge.

We will use this possibility to control the errors arising from the
numerical integration of the perturbation equations. We will express
both our initial conditions and the final results for the perturbation
variables in the ZSG, and use the UCG simply as an estimator of the
numerical integration errors. As an example, we will derive the
relations between the perturbations in the energy density of the ideal
fluid in the ZSG and the UCG. The change from the ZSG to the UCG, and
vice-versa, for the other perturbation variables can be obtained in a
similar way.

The quantities 
\begin{eqnarray}
\label{gaugeinvZSG}
\epsilon_{\gamma}^{\chi} & \equiv & \epsilon_{\gamma}+
	3H(\mu_{\gamma}+p_{\gamma})\chi\,,\\
	\Psi_{\gamma}^{\chi} & \equiv & \Psi_{\gamma}+
	(\mu_{\gamma}+p_{\gamma})\chi\,,
\end{eqnarray}
and 
\begin{eqnarray}
\label{gaugeinvlUCG}
\epsilon_{\gamma}^{\varphi}& \equiv & \epsilon_{\gamma}+
	3(\mu_{\gamma}+p_{\gamma})\varphi\,,\\
	\Psi_{\gamma}^{\varphi} & \equiv & \Psi_{\gamma}+
	(\mu_{\gamma}+p_{\gamma})\frac{\varphi}{H}\,,
\end{eqnarray}
are invariant under temporal gauge transformations, as can be seen by
using the relations provided by Hwang \cite{Hw2}. The first two
variables simply take the values of $\epsilon_{\gamma}$ and
$\Psi_{\gamma}$, respectively, when these quantities are calculated in
the ZSG, while the same occurs for the last two variables with
relation to the UCG. We thus want to express in the UCG
$\epsilon_{\gamma}^{\varphi}$ as a function of
$\epsilon_{\gamma}^{\chi}$ and $\Psi_{\gamma}^{\chi}$, in order to
obtain the initial value of $\epsilon_{\gamma}$ to be used in the UCG
calculations from that originally given in the ZSG. Also, we want to
know how to obtain $\epsilon_{\gamma}^{\chi}$ from
$\epsilon_{\gamma}^{\varphi}$ and $\Psi_{\gamma}^{\varphi}$, so that
we can compare the final value of $\epsilon_{\gamma}$ obtained in the
two gauges.

In the UCG we have 
\begin{equation}
\label{grelUCG1}
\epsilon_{\gamma}^{\varphi}=\epsilon_{\gamma}^{\chi}-
	3H(\mu_{\gamma}+p_{\gamma})\chi\,,
\end{equation}
and, by using Eqs.~(\ref{pert2}) and (\ref{pert3}), 
\begin{equation}
\label{grelUCG2}
\frac{k^{2}}{a^{2}}H\chi-12\pi GH(\Psi_{\gamma}+\Psi_{\phi})=
	-4\pi G(\epsilon_{\gamma}+\epsilon_{\phi})\,.
\end{equation}
Through some algebraic manipulation of the above relations we then obtain 
\begin{equation}
\label{grelUCG3}
\epsilon_{\gamma}^{\varphi}=\epsilon_{\gamma}^{\chi}-
	12\pi G(\mu_{\gamma}+p_{\gamma})\frac{a^{2}}{k^{2}}
	\left[3H(\Psi_{\gamma}^{\chi}+\Psi_{\phi}^{\chi})-
	(\epsilon_{\gamma}^{\chi}+\epsilon_{\phi}^{\chi})\right]\,, 
\end{equation}
where $\epsilon_{\phi}^{\chi}$ and $\Psi_{\phi}^{\chi}$ are defined in
an analogous fashion to the ideal fluid gauge-invariant variables.

On the other hand, in the ZSG we have 
\begin{equation}
\label{grelZSG1}
\epsilon_{\gamma}^{\chi}=\epsilon_{\gamma}^{\varphi}-
	3(\mu_{\gamma}+p_{\gamma})\varphi\,,
\end{equation}
and, by using the same equations as previously, 
\begin{equation}
\label{grelZSG2}
-\frac{k^{2}}{a^{2}}\varphi-12\pi GH(\Psi_{\gamma}+\Psi_{\phi})=
	-4\pi G(\epsilon_{\gamma}+\epsilon_{\phi})\,.
\end{equation}
Hence, we get 
\begin{equation}
\label{grelZSG3}
\epsilon_{\gamma}^{\chi}=\epsilon_{\gamma}^{\varphi}-
	12\pi G(\mu_{\gamma}+p_{\gamma})\frac{a^{2}}{k^{2}}
	\left[(\epsilon_{\gamma}^{\varphi}+\epsilon_{\phi}^{\varphi})-
	3H(\Psi_{\gamma}^{\varphi}+\Psi_{\phi}^{\varphi})\right]\,, 
\end{equation}
where again $\epsilon_{\phi}^{\varphi}$ and $\Psi_{\phi}^{\varphi}$
are defined in the same way as the ideal fluid gauge-invariant
variables.


\end{document}